\documentstyle[12pt]{article}

\begin{document}
\begin{flushright}
{BI-TP 94/40}
\end{flushright}
\begin{flushright}
{July 1994}
\end{flushright}
\begin{center}
{\bf ON THE DECONFINEMENT PHASE TRANSITION IN THE RESONANCE GAS}
\end{center}
\begin{center}
{\bf A.V. Leonidov\footnote{Alexander von Humboldt Fellow}$^{(a,b)}$
and G.M.Zinovjev $^{(c)}$}
\end{center}
\begin{center}
{\it (a) Physics Department\\ University of Bielefeld 33615
Bielefeld,Germany}\\
\end{center}
\begin{center}
{\it (b) Theoretical Physics Department\\ P.N.Lebedev
Physics Institute, 117924 Leninsky pr.53, Moscow, Russia}\\
\end{center}
\begin{center}
{\it (c) N.N.Bogolyubov Institute for Theoretical Physics\\ 252143 Kiev,
Ukraine}\\
\end{center}
\begin{abstract}
We obtain the constraints on the ruling
parameters of the dense hadronic gas model
at the critical temperature and propose the quasiuniversal
ratios of the
thermodynamic quantities.
The possible appearence of thermodynamical instability
is discussed.
\end{abstract}
\newpage

{\bf{1.}} The thermodynamics of a dense hadron gas is bound to be
a focal point of the analysis of possible phase transformations in the
hadronic matter and their manifestations.
With the energies of the heavy ion colisions available at present and,
most importantly, in the near future the topic is no longer a purely
theoretical one. In high energy nuclear collisions
we observe the thermodynamics of strong interactions in action [1].
If we imagine the
initially dilute hadronic gas being heated, then at some point the
hadronic excitations are overlapping to an extent that does not any
longer permit
to use the picture of a weakly interacting hadron gas.  In this context
the most
reliable estimate of a temperature at which the hadron gas density is
high enough and this description is no longer true is
provided by a chiral theory [2] and gives a value of $T'=130 \mbox { MeV
 }$. At temperatures $T > m_{\pi}$ we have a strong interacting
dense hadronic system. As we believe that at high enough temperatures the
true excitations are gluons and quarks, the crucial questions are at which
temperature and how the hadron phase is transforming into a quark-gluon one.
Indeed, it is highly unprobable that the transition takes place within a weakly
coupled hadronic regime (i.e. at $T < T'$), and we are necessarily facing a
problem of describing a strong coupled regime of a dense hot hadron gas.
An ingenious way of dealing with this obviously desperate situation was
proposed by Hagedorn (see, e.g., [3]). The idea was to rewrite a
partition function of a strongly interacting system as that of an
infinite number of resonances having zero width and characterized by a
rapidly (exponentially) growing spectrum thus diagonalizing the
initial hamiltonian {\footnote{This idea was later succsessfully
realized for two-dimensional conformal theories and integrable
deformations of them (see, e.g., [4])}}. The resulting hadronic mass
spectrum  takes a form
\begin{equation}
\tau(m)=c {1 \over m^a} exp({m \over T_0})
\end{equation}
depending on three parameters $a,c$ and $T_0$. The typical expression
for the thermodynamic quantity (here at the example of a pressure of a
relativistic gas of massive particles) has a form
\begin{equation}
P= \sum_i {m_i^2 T^2 \over 2 \pi^2} K_2({m_i \over T})+
{T^2 \over 2\pi^2}\int_{M}^{\infty}dm \tau(m)m^2K_2({m \over T})
\end{equation}
where the sum is taken over the discrete spectrum $(m_i < M)$ and the
integration is performed over the continuous one.
We emphasise that the fourth parameter has appeared: the cutoff mass $M$.
Below we argue that it can in fact play a decisive role in the
description of the hadronic phase.

{\bf{2.}} Now when the framework of the discussion and the set of
parameters of the hadronic phase is settled, let us briefly remind the
existing possibilities for arranging the transition between the hadronic
matter and the quark-gluon one. The key parameter is a power in the
preexponential mass factor in Eq.(1)
that determines the type of a possible
singularity of a partition function. In particular, for $a \leq  7/2$ the
energy density becomes infinite at $T=T_0$, when $a=9/2$ the energy
density is finite at $T=T_0$, but the specific heat has a logarithmic
singularity similar to that in the second order phase transitions,
and at $a>9/2$ the singularity moves to higher orders in the
derivatives of the thermodynamic potentials (see, e.g., [3,5,6]).
A more sophisticated treatment of hadron bag gas  within alternative
pressure ensemble leads  to similar conclusions [7].

Now let us discuss at which temperature $T_c$ the transition to the
quark-gluon phase takes place. The physical situation will obviously
depend on the value of the parameter $a$.  Before listing the arising
possibilities let us remind that the calculation of the thermodynamical
quantities using the exponential spectrum (1) is far from being trivial.
In a seminal paper [8] Carlitz has shown that for $a>5/2$ the
equivalence of the canonical and microcanonical ensembles is no longer
guaranteed and the system can be thermodynamically unstable (negative
specific heat) thus making the statistical approach questionable.
Below we give more comments on that.

In order to get a quantitative estimate of the possible value of the
transition temperature $T_c$ it is customary to equate the expressions
for the pressure in the hadronic phase (with some given value of $a$) to
that in the quark-gluon gas, for which one usually takes the
bag-motivated expression
\begin{equation}
P_{QGP}(T)={37 \pi^2 \over 90}T^4-B_{0}
\end{equation}
 where $B_{0}$ is a nonperturbative vacuum pressure (bag constant).

At $a \leq 7/2$ , when $T_0$ is a limiting temperature, there is no
crossover of the pressure
curves (2) and (3) (see, e.g., [9]). This have led the
authors of several papers [10] to the necessity of introducing additional
mass-dependent interactions eliminating the limiting temperature.
However, this proposal contradicts the original idea, namely
that of hiding all the interaction effects in the infinite spectrum of
{\it free} resonances. Therefore it is interesting to look more closely at
other possibilities of arranging a phase transition working only with a
free spectrum (1).

{\bf{3.}}  The obvious attempt to handle this unfortunate situation
is to try to construct a
phase transition choosing $a > 9/2$, so that the thermodynamic
potentials and their derivatives up to the second order (including the
specific heat) are finite at $T=T_0$.

Before turning to the pressure equality equations let us stress again [8],
that for $a > 9/2$ we find ourselves in a domain where the thermodynamical
stability of a system can break down. In particular, the
specific heat is positive for the energy densities
\begin{equation}
\varepsilon < \varepsilon_{crit} = {M \over V}
\end{equation}
where $V$ is a volume of a system and $M$ is a cut-off used in
integrating over the continous spectrum (see Eq. (2)). Let us now imagine
the evolution of the quark-gluon matter created in heavy ion collisions
or in the early Universe.
If
$$
\varepsilon_{QGP}(T_c) > \varepsilon_{crit}
$$
the produced hadron matter will form a thermodynamically unstable
system with the preferable configuration consisting of the particles
all having masses of order of $M$. This could lead to interesting
predictions for the heavy ion collisions. Namely that can give rise to a
explosive production of particles with the masses of order of
cut-off one.
Here we would like to stress the crucial role of the cut-off mass
$M$. The value of this cut-off is usually beleived to be a matter of
choice, because there is no natural border between the discrete and
continuum spectrum. However as there are good grounds to believe that
the masses in the discrete part of the spectrum (pions, nucleons, ...)
significantly depend on
the interaction of the corresponding particles (e.g., decrease with
temperature), the account for interactions in the discrete spectrum
forces us, generally speaking, to make corresponding changes in the
value of the cut-off mass $M$. For example, if, following the "mean
field" logic, the masses of the particles are decreasing with increasing
density, thus forcing us to decrease the value of $M$, then because of
the dropping $\varepsilon_{crit}$ we are increasing the probability of
getting a thermodynamically unstable system. This is also potentially
important for determining the temperature $T'$. The necessity of
shifting the cut-off mass $M$ towards lower values can invalidate the
picture of a dilute hadron gas earlier than naively expected.

{\bf 4.} Let us now suppose, that $\varepsilon_{crit}$ is big enough so that
the
system is thermodynamically stable. Then one can carry out a standard
Maxwell construction
getting the intersection of the pressure curves for
both phases and thus a phase transition point.
This is of course an assumption, but at $c_v>0$ one can believe that the
microcanonical and canonical ensembles are equivalent [11].
In the following we shall
exploit a simplest hypothesis, namely that a transition takes place at
$T=T_0=T_c$.
The densities of the thermodynamical
quantities at $T_c$ can be calculated analytically. Let us consider
$T_c<M$ (which holds for any realistic description of the phase
transition in the hadronic matter). The contribution of the
discrete spectrum at $T_c$ is exponentially smaller than that of the
continuum one and we get for the presssure and entropy
\begin{equation}
P(T_c)={cT_{c}^{5-a} \over (2\pi)^{3/2}}\sum_{n=0}
{(2,n) \over 2^n}{(T_c/M)^{a-5/2+n} \over a-5/2+n}
\end{equation}
\begin{equation}
S(T_c)={cT_{c}^{4-a} \over (2\pi)^{3/2}}\sum_{n=0}
{(3,n) \over 2^n}{(T_c/M)^{a-7/2+n} \over a-7/2+n}
\end{equation}
where $(\nu,n)$ is a Hankel's symbol
$$
(\nu,n)={\Gamma(1/2+\nu+n) \over n! \Gamma(1/2+\nu-n)}
$$
The conditions for
the deconfinement transition to take place are (all quantities are taken
in the critical point):
\begin{equation}
\cases{P_{h}=P_{QGP}>0\cr S_{h}<S_{QGP}\cr},
\end{equation}
where $S_{h,QGP}$ are the entropy densities in the hadron and
quark-gluon phases respectively.
Working in the lowest order in the small parameter $\rho=T_{c}/M$ and
neglecting the contribution of the discrete spectrum  we get the
following restrictions on the values of the parameters present in the
hadron and plasma partition functions:
\begin{equation}
({37\pi^2 \over 90})^{1/4}(1-4{1+\varepsilon \over
2+\varepsilon}\rho)^{1/4}T_{c}<B_{0}^{1/4}<({37\pi^2 \over 90})^{1/4} T_{c}
\end{equation}
and
\begin{equation}
c<(2\pi)^{3/2}{74\pi^2 \over 45} T_{c}^{7/2+\varepsilon} (1+\varepsilon)
\rho^{1+\varepsilon}
\end{equation}
where,as discussed before, $\varepsilon=a-9/2>0$. As the bag constant $B_{0}$
should be positive, one also gets from Eq. (9):
\begin{equation}
{T_{c} \over M}<{2+\varepsilon \over 4(1+\varepsilon)}
\end{equation}
The corresponding numerical restrictions on
the possible values of the bag constant $B_{0}$ and the parameter $c$
for some typical values of the dimensionful parameters $T_{0}$ and $M$
and $a=5$ are given in the Table.  Thus we see that the possibility of
a phase transition heavily restricts the values of the parameters in the
problem.  This is especially clear for the values of the
parameter $c$ which are quite  small. Let us notice
however that phenomenologically there is practically no restrictions on
the values of parameters in the spectrum (1), because this formula is
supposed to be valid only for the highly excited resonances where the
hadron spectrum is pretty unknown. In this respect the small allowed
values of $c$ are intuitively uncomfortable but not phenomenologically
forbidden.

Let us note, that the structure of the expressions for the thermodynamical
quantities Eqs. (5), (6)  enables one to get the information about the phase
transition which is to a large extent independent of the parameters
chosen to describe a hadronic phase near the first order phase
transition point. The first example of such a quasiuniversal ratio is
\begin{equation}
{P(T_c) \over E(T_c)}=({a-7/2 \over a-5/2}){T_c \over M}
\end{equation}
where we are using the lowest order expressions in the small parameter
$T_{c}/M$. As in order to describe a first order phase transition
it is necessary to have $a>9/2$, we get one more inequality for
$T_{c}/M$:
\begin{equation}
{ 1 \over 2}{P(T_c) \over E(T_c)}<{T_c \over M}<{P(T_c) \over E(T_c)}
\end{equation}
Let us also calculate the derivative of the energy over temperature
(as the calculations are done exclusively within the hadronic phase,
the derivative should be understood as a left one). In the leading order
in $T_c/M$ we have
\begin{equation}
\left.{\partial E \over \partial T}\right|_{T_c}=cT_{c}^{4-a}{1 \over
a-9/2} ({T_c \over M})^{a-9/2}
\end{equation}
Now we can form another interesting ratio
\begin{equation}
{P(T_c)T_cE'(T_c) \over E^{2}(T_c)}=1+{1 \over (a-9/2)^2+2(a-9/2)}
\end{equation}
which for $a>9/2$ is always larger than one. Thus we see that the
proposed description of the first order phase transition leads to some
quasiuniversal ratios (weakly dependent only on the parameter $a$ and
independent of all other parameters)
which are in some sense analogous to the universality in the
description of the second order phase transition.
In the picture we are
discussing this is not suprising, because the exponentially growing
spectrum leads to a powerlike temperature dependence
in both situations, and the difference depends on the value of the
parameter $a$.
These restrictions seem to be fairly important for
the calculations
describing the matter evolution in high energy nuclear collisions and
early Universe.

We are grateful to H.Satz for useful discussions.

\begin{center}
{\it References}
\end{center}

 1. H.Satz, {\it{Nucl.Phys.}} {\bf{A566}} (1994), 1c;

 2. P.Gerber and H.Leutwyler, {\it{Nucl.Phys.}} {\bf{B321}} (1989), 387;

 3. R.Hagedorn, {\it{Nuov.Cim.Suppl.}} {\bf{3}} (1965), 117;

 4. Vl.S.Dotsenko, {\it{Adv.Studies in Pure Math.}} {\bf{16}} (1987),
 123;
 G.Mussardo, {\it{Phys. Rep.}} {\bf{C218}} (1992), 215;

 5. K.Huang and S.Weinberg, {\it{Phys.Rev.Lett.}} {\bf{25}} (1970), 895;

 6. H. Satz, {\it{Phys.Rev.}} {\bf{D19}} (1979), 1912;

 7. M.I.Gorenstein et al., {\it{Theor.Math.Phys.}} {\bf{52}} (1982),
 843;

 8. R.D.Carlitz, {\it{Phys.Rev.}} {\bf{D5}} (1972), 3231;

 9. J.I.Kapusta, "Finite Temperature Field Theory", Cambridge University
 Press, 1989;

 10.  K.A.Olive, {\it{Nucl.Phys}} {\bf{B190[FS3]}} (1981), 483;
    K.A.Olive, {\it{Nucl.Phys}} {\bf{B198}} (1982), 461;
    J.I.Kapusta and K.A.Olive, {\it{Nucl.Phys}} {\bf{A408}} (1983), 478;
    B.A.Campbell, J.Ellis and K.A.Olive, {\it{Nucl.Phys}} {\bf{B345}} (1990),
 57;
    R.Venugopalan and M.Prakash, {\it{Nucl.Phys}} {\bf{A546}} (1992),
 718;

 11. H.A.Weldon, {\it {Journ.of Phys.}} {\bf {193}} (1989), 195.

\begin{center}
 Table
\end{center}

\begin{tabular}{|c|c|c|c|c|}   \hline
$M$ (GeV) &$T_c$ (GeV) & $B_{0 min}^{1/4}$ (GeV)& $B_{0 max}^{1/4}$ (GeV)
&$c_{max}$ \\ \hline
1 &0.16& 0.20& 0.23& 0.02 \\ \hline
1 &0.2 &0.24 &0.28 &0.05 \\ \hline
1 &0.24 &0.27 &0.34 &0.15 \\ \hline
2 &0.16 &0.21 &0.23 &0.01 \\ \hline
2 &0.2 &0.26 &0.28 &0.02 \\ \hline
2 &0.24 &0.31 &0.34 &0.05 \\ \hline
\end{tabular}

\begin{center}
{\it Table Caption}
\end{center}

 Restrictions on model parameters at $\varepsilon=a-9/2=0.5$.

\end{document}